\documentclass[12pt]{article}

\long\def\symbolfootnote[#1]#2{\begingroup%
\def\thefootnote{\fnsymbol{footnote}}\footnote[#1]{#2}\endgroup}

 \font\tenrm=cmr10 \font\tenit=cmti10
\font\elevenbf=cmbx10 scaled\magstep 1 
scaled\magstep 1  1

\newcommand{\x}{\hat{x}}
\newcommand{\be}{\begin{equation}}
\newcommand{\bea}{\begin{eqnarray}}
\newcommand{\eea}{\end{eqnarray}}
\newcommand{\non}{\nonumber}
\newcommand{\ee}{\end{equation}}
\newcommand{\la}{\label}

\newcommand{\oal}{\overline{\alpha}}
\newcommand{\opi}{\overline{\pi}}

\newcommand{\D}{\mbox {D}}
\newcommand{\omu}{\overline{\mu}}

\newcommand{\onu}{\overline{\nu}}
\newcommand{\obe}{\overline{\beta}}

\newcommand{\oga}{\overline{\gamma}}

\newcommand{\ode}{\overline{\delta}}
\newcommand{\oio}{\overline{\iota}}

\newcommand{\oo}{\overline{o}}
\newcommand{\dA}{\dot{A}}
\newcommand{\dB}{\dot{B}}
\newcommand{\dC}{\dot{C}}

\newcommand{\dD}{\dot{D}}

\newcommand{\dE}{\dot{E}}

\newcommand{\dF}{\dot{F}}

\newcommand{\ola}{\overline{\lambda}}
\newcommand{\al}{\alpha}
\newcommand{\de}{\delta}

\def\s={\stackrel{*}{=}}
\def\g0{\stackrel{\circ}{g}}

\def\x{x_1}
\def\y{x_2}
\def\xc{\overline{x_1}}
\def\yc{\overline{x_2}}

\def\be{\begin{equation}}
\def\bea{\begin{eqnarray}}

\def\ee{\end{equation}}
\def\eea{\end{eqnarray}}

\def\square{\vcenter{\vbox{\hrule height.5pt \hbox{\vrule
width.5ptheight7pt \kern7pt \vrule width.5pt} \hrule height.5pt}}}

\begin{document}
\newtheorem{theorem}{Theorem}
\newtheorem{lemma}{Lemma}

\begin{center}
\vglue 0.6cm

 {\elevenbf        \vglue 10pt
    Complete Solution of Hadamard's Problem for the Scalar Wave
Equation  on Petrov type III Space-Times
\symbolfootnote[2]{Published
on Ann. Inst. Henri Poincar\'e (A) Phys. Th\'eorique, \textbf{71}, 595 (1999).}}\\
\vglue 0.5cm

{\tenrm S. R. Czapor\\} \baselineskip=13pt {\tenit Department of
Mathematics and Computer Science, Laurentian University \\}
\baselineskip=12pt
{\tenit Sudbury, Ontario, Canada P3E 2C6}\\
\vglue 0.5cm {\tenrm R. G. McLenaghan\\} \baselineskip=13pt
{\tenit Department of Applied Mathematics, University of Waterloo
\\} \baselineskip=12pt
{\tenit Waterloo, Ontario, Canada N2L 3G1}\\
\vglue 0.5cm
{\tenrm and }\\
\vglue 0.5cm {\tenrm F. D.
Sasse\symbolfootnote[1]{fsasse@joinville.udesc.br}\\}
\baselineskip=12pt {\tenit Department of Mathematics, Centre for
Technological Sciences-UDESC
\\} \baselineskip=12pt {\tenit Joinville 89223-100, Santa
Catarina, Brazil} \vglue 1.0cm { \tenrm\baselineskip=12pt
 \noindent
\begin{quote}
ABSTRACT - We prove that there are no Petrov type III space-times
on which the conformally invariant (self-adjoint) scalar wave
equation or the non-self-adjoint scalar wave equation satisfies
Huygens' principle.
\end{quote}
} \vglue 0.8cm { \tenrm\baselineskip=12pt
 \noindent
\begin{quote}
R\'{E}SUM\'{E} - Nous prouvons qu'il n'existe aucun espace-temps
de type III de Petrov sur lequel l'\'{e}quation invariante
conforme des ondes scalaires ou l'\'{e}quation des ondes scalaires
non-auto-adjoint satisfait au principe de Huygens.
\end{quote}
}
\end{center}

\section{Introduction}
\la{intr} This paper is  devoted to the the solution of Hadamard's
problem on Petrov type III space-times, for the conformally
invariant scalar wave equation \be \la{selfad} \square\,u
+\frac{1}{6}Ru=0\,, \ee and the non-self-adjoint scalar wave
equation \be \la{nsa} \square\,u + A^a \partial_a u + Cu=0\,. \ee
In the above equations $\,\square\,$ denotes the Laplace-Beltrami
operator corresponding to the metric $g_{ab}$ of the background
space-time V${}_4$, $u$ the unknown function, $R$ the Ricci
scalar, $A^a$ the components of a given contravariant vector field
and $C$ a given scalar function. The background manifold, metric
tensor, vector field and scalar function are assumed to be
$C^{\infty}$. All considerations of this paper are entirely local.

The homogeneous equations (\ref{selfad}) and (\ref{nsa}) satisfy
{\it Huygens' principle} in the sense of Hadamard \cite{Had23} if
$u(x)$ depends only on the Cauchy data in an arbitrarily small
neighborhood of the intersection between the backward
characteristic conoid  $C^-(x)$ with the vertex at $x$ and the
initial surface $S$, for arbitrary Cauchy data on $S$, arbitrary
$S$, and for all points $x$ in the future of $S$. {\it Hadamard's
problem} for (\ref{selfad}) and (\ref{nsa}) is that of determining
all space-times for which Huygens' principle is valid. We recall
that two equations (\ref{nsa}) are said to be {\it equivalent} if
and only if one may be transformed into the other by any
combination of the following {\it trivial transformations}:
\newline
$\,$
\newline
a) a general coordinate transformation;
\newline
b) multiplication of the equation by the function
$exp(-2\phi(x))$, which induces a {\it conformal transformation}
of the metric
\begin {eqnarray}
\nonumber {\tilde{g}}_{ab} = e^{2\phi}g_{ab} \,;
\end {eqnarray}
\newline
c) substitution of $\lambda u$ for the unknown function $u$, where
$\lambda$ is a non-vanishing function on V${}_4$.
\newline
$\,$
\newline
We note that the Huygens' character of (\ref{nsa}) is preserved by
any trivial transformation. In the case of (\ref{selfad}) the
trivial transformations reduce to conformal transformations with
$\lambda = e^\phi$.

Carminati and McLenaghan \cite{car82} have outlined a program for
the solution of Hadamard's problem for the scalar wave equation,
Weyls' neutrino equation and Maxwell's equations based on the
conformally invariant Petrov classification of the Weyl conformal
curvature tensor. This involves the consideration of five disjoint
cases which exhausts all the possibilities for non-conformally
flat space-times. Hadamard's problem for (\ref{selfad}) and
(\ref{nsa}) has been completely solved for Petrov type N
space-times by Carminati and McLenaghan \cite{car84,car86} and
McLenaghan and Walton \cite{mcl88}. Their results may be
summarized as follows:
\begin{quote}{\it
Any non-self-adjoint equation (\ref{nsa}) on any Petrov type N
background space-time satisfies Huygens' principle if and only if
it is equivalent to the wave equation $\square \,u = 0$ on an
exact plane wave space-time with metric }\end{quote}
\be \label{planew} ds2 = 2dv\{du + [D(v)z2 +
\overline{D}(v)\overline{z}^2 + e(v)z\overline{z}]dv\} -
2dzd\overline{z} \,. \ee

For Petrov type D space-times the following result was obtained by
Carminati and McLenaghan \cite{car87}, McLenaghan and Williams
\cite{mcl90} and W\"unsch \cite{wun89}:
\begin{quote}{\it
 There exist no Petrov type D space-times on which the
conformally invariant scalar wave equation (\ref{selfad})
satisfies Huygens' principle. }\end{quote} In the present paper we
complete this program for the conformally invariant scalar wave
equation (\ref{selfad}) on Petrov type III space-times by proving
the following theorem:
\begin{theorem}
\la{teo-main} There exist no Petrov type III space-times on which
the conformally invariant scalar wave equation (\ref{selfad})
satisfies Huygens' principle.
\end{theorem}
The results on type N and type D space-times described above and
Theorem 1 lend weight to the conjecture which states that every
space-time on which the conformally invariant scalar wave equation
satisfies Huygens' principle is conformally related to the plane
wave space-time (\ref{planew}) or is conformally flat
\cite{car84,car86}.

Hadamard's problem for the general non-self-adjoint equation
(\ref{nsa}) may now be solved with the help of Theorem
\ref{teo-main} and the results of Anderson, McLenaghan and Sasse
\cite{and99} where the following theorem is proved:
\begin{theorem}
\la{teo-nsa} Any non-self-adjoint scalar wave equation (\ref{nsa})
which satisfies Huygens' principle on any Petrov type III
background space-time is equivalent to the conformally invariant
scalar wave equation (\ref{selfad}).
\end{theorem}
Combining these two theorems we obtain
\begin{theorem}
\la{teo-fin} There exist no Petrov type III space-times on which
the non-self-adjoint scalar wave equation satisfies Huygens'
principle.
\end{theorem}
The corresponding problem for the Weyl neutrino equation and
Maxwell's equations is solved in \cite{mcl96}.

The starting point of our proof of Theorem \ref{teo-main} is the
paper by  Carminati and McLenaghan \cite{car88}, where the
following results are obtained for Petrov type III space-times:
\begin{theorem}
\la{teo1} The validity of Huygens' principle for the conformally
invariant scalar wave equation (\ref{selfad}), on any Petrov type
III space-time implies that the space-time is conformally related
to one in which every repeated principal spinor field $o_A$ of the
Weyl spinor is recurrent, that is \be o_{A;B \dot{B}}=o_AI_{B
\dot{B}}\,, \ee where $ I_{B \dot{B}} $ is a 2-spinor, and \be
\Psi_{ABCD;E \dot{E}}\,\iota^A \iota^B \iota^C o^D o^E
\overline{o}^{\dot{E} }=0\,, \ee \be R=0\,,\;\;\;\;\;\;\Phi_{AB
\dot{A} \dot{B}}o^A o^B=0\,, \ee where $\iota^A$ is any spinor
field satisfying $o_A \iota^A = 1$.
\end{theorem}
\begin{theorem} If any one of the following three conditions
\la{teo2} \be \la{eq1teo2} \Psi_{ABCD;E \dot{E}}\, \iota^A \iota^B
\iota^D \iota^E \overline{o}^ {\dot{E}}=0\,, \ee \be \la{eq2teo2}
\Psi_{ABCD;E \dot{E}}\, \iota^A \iota^B o^D o^E \overline{\iota}^
{\dot{E}}=0\,, \ee \be \la{eq3teo2} \Psi_{ABCD;E \dot{E}}\,
\iota^A \iota^B \iota ^D o^E \overline{o}^ {\dot{E}}=0\,, \ee
 is satisfied, then there exist no Petrov type III space-times on which
the conformally invariant scalar wave equation (\ref{selfad})
  satisfies Huygens' principle.
\end{theorem}
It is important to note that these earlier results solve
Hadamard's problem under what have proved to be fairly strong
assumptions (namely, that one of (\ref{eq1teo2}), (\ref{eq2teo2}),
or (\ref{eq3teo2}) is satisfied). The purpose of the present paper
is to make the analysis completely general by removing these
assumptions. We follow the conventions of \cite{car88}, and use
the results established there to obtain (most of) the basic
equations needed for the proof of Theorem \ref{teo-main}.

In Section \ref{sec-2} we give the necessary conditions for the
validity of Huygens' principle that will be used in this paper,
and give a brief summary of their implications.  From these
necessary conditions, we derive the further side relations needed
for our analysis in terms of the Newman-Penrose scalars. The key
to our proof is the six-index necessary condition obtained by
Rinke and W\"unsch \cite{rink81} which was not used in
\cite{car88}. In Section \ref{sec-3} we examine these side
relations in the case $\Phi_{11}=0$ and show that they lead to a
contradiction. The proof of Theorem \ref{teo-main} is completed in
Section \ref{sec-4}, where the case $\Phi_{11} \neq 0$ is treated.

It is worth mentioning that the tools of computer algebra are used
throughout this paper. Initially we employ  Maple  package
NPspinor \cite{cza87,cza92} to extract dyad components of spinor
versions of the necessary conditions, and then to manipulate the
resulting expressions in Newman-Penrose form. In the case
$\Phi_{11}=0$ we use the Gr\"{o}bner basis package of the Maple
system to explicitly determine solutions of systems of algebraic
equations. Finally, for the case $\Phi_{11} \neq 0$ we use the GB
package of Faug\`ere \cite{fau94} to examine the solvability of a
somewhat larger system of algebraic equations.

\section{Formalism and Basic Equations}
\la{sec-2} The necessary conditions for the validity of Huygens'
principle for (\ref{selfad}) which we employ are given by \be
\label{conIII} (III) \qquad S_{abk;}{}^{k}\,-\,
\frac{1}{2}C^{k}{}_{ab}{}^{l}L_{kl}=0\,, \ee
\begin{eqnarray}
\nonumber (V)\;&& TS\left(
3C^k{}_{ab}{}^l{}_{;}{}^mC_{kcdl;m}\,+\,8C^k{}_{ab}{}^l{}_{;c}
S_{kld}+40S_{ab}{}^k{}S_{cdk}-8C^k{}_{ab}{}^lS_{klc;d} \right.\\
\label{conV} &&\left.-24C^k{}_{ab}{}^lS_{cdk;l}\,+\,
4C^k{}_{ab}{}^lC_l{}^m{}_{ck}L_{dm}+12C^k{}_{ab}{}^lC^m{}_{cdl}L_{km}\right)=0\,,
\end{eqnarray}
\be \label{conVII} (VII)\quad
TS\left(Q^{(1)}_{abcdef}-10Q^{(2)}_{abcdef}+4Q^{(3)}_{abcdef}
+5Q^{(4)}_{abcdef}+Q^{(5)}_{abcdef}\right)=0\,, \ee where
\begin{eqnarray}
\nonumber
Q^{(1)}_{abcdef}&=&3C^k{}_{ab}{}^l{}_{;}{}^m{}_{c}C_{kdel;mf}+
C^k{}_{ab}{}^l{}_{;cd}(10S_{kle;f}+6S_{efk;l})
+64S_{abk;c}S_{de}{}^k{}_{;f}\\
\nonumber &&-C^k{}_{ab}{}^l\left(
3C^m{}_{cdk;ef}L_{lm}+5C_{kcdl;me}L^m{}_f+
7C^m{}_{cdk;le}L_{mf}\right.\\
\label{2.193}
&&\left.+13S_{klc;d}L_{ef}+12S_{cdk;l}L_{ef}+71S_{cdk;e}L_{lf}
\right)\,,
\end{eqnarray}
\begin{eqnarray}
\nonumber
Q^{(2)}_{abcdef}&=&C^k{}_{ab}{}^l{}_{;c}\left(S_{kld;ef}+3S_{dek;lf}
+2S_{abk;cd}S_{ef}{}^k-5S_{abk}S_{cd}L_{ef}\right)\\
\nonumber &&-\frac{1}{2}C^k{}_{ab}{}^l{}_{;c}\left( 2
C^m{}_{kld;e}L_{mf}+3C^m{}_{dek;l}L_{mf}+S_{kld}L_{ef} \right.\\
\nonumber
&&\left.+3C_{kde}{}^m{}_{;f}L_{lm}+15S_{dek}S_{lf}\right)-
C^k{}_{ab}{}^l\left(C_{kcd}{}^m{}_{;e}L_{(lm;f)}\right.\\
\label{2.194} &&+S_{cdk}L_{(le;f)}-\frac{1}{12}R_{;c}C_{kdel;f}
)\,,
\end{eqnarray}
\begin{eqnarray}
\nonumber
Q^{(3)}_{abcdef}&=&-C^k{}_{ab}{}^l\left(2C_k{}^{mn}{}_cC_{lnmd;ef}-
10C^m{}_{cd}^nC_{kefl;mn}+20C_{lcd}{}^mS_{kme;f}\right)\\
\nonumber &&-5C_k{}^{mn}{}_aC_{lmnb}
C^k{}_{cd}{}^l{}_{;ef}+C^k{}_{ab}{}^l \left(7C_k{}^{mn}{}_c
C_{lmnd}L_{ef}\right.\\
\label{2.195} &&\left.-10C_{kefl}C^m{}_{cd}{}^nL_{mn}\right)\,,
\end{eqnarray}
\begin{eqnarray}
\nonumber
Q^{(4)}_{abcdef}&=&-C^k{}_{ab}{}^l\left(2C_k{}^{mn}{}_{c;d}C_{lmnd;ef}+
54C_{lcd}{}^m{}_{;e}S_{kmf}+74C_{lcd}{}^m{}_{;k}S_{efm}\right.\\
\label{2.196} &&- \frac{76}{3}C_{ckl}{}^m{}_{;d}S_{efm}-
\frac{404}{3}S_{cdk}S_{efl})  +6C_k{}^{mn}{}_aC^k{}_{bc}{}_{;d}
C_{lefm;n}\,,
\end{eqnarray}
\begin{eqnarray}
\label{2.197} Q^{(5)}_{abcdef}=-C^k{}_{ab}{}^l
C_{lcd}{}^mL_{km}L_{ef}
+\frac{1}{6}C^k{}_{ab}{}^lC_{kcdl}(87L^m{}_{e}L_{mf}
+19RL_{ef})\,.
\end{eqnarray}
where
\begin{eqnarray}
\la{weyl}
&&C_{abcd}:=R_{abcd}-2g_{[a[d}L_{b]c]}\,,\\
\la{S}
&&S_{abc}:=L_{a[b;c]}\,,\\
\la{L} &&L_{ab}:=-R_{ab}+\frac{R}{6} g_{ab}\,.
\end{eqnarray}
Here $A_a:=g_{ab}A^b$, $ R_{abcd}$ denotes the Riemann tensor,
$C_{abcd}$ the Weyl tensor, $R_{ab}:= g^{cd}R_{cabd}$, the Ricci
tensor, and $R:=g^{ab}R_{ab}$ the Ricci scalar associated to the
metric $g_{ab}$. The conditions $III$, $V$ and $VII$ are
necessarily conformally invariant. Spinor versions of conditions
$III$ and $V$, and the conventions used for conversion from the
original tensor form, are given in \cite{car88}.

Mathisson \cite{mat39}, Hadamard \cite{Had42}, and Asgeirson
\cite{asg56} obtained condition $III$ for (\ref{nsa}) in the case
$g^{ij}$ constant. Condition $III$ was obtained in the general
case for (\ref{nsa}) by G\"unther \cite{gun52}. Condition $V$ was
obtained by McLenaghan \cite{mcl69} in the case $R_{ab}=0$, and by
W\"unsch \cite{wun70} for the general case.  Condition $VII$ was
obtained by Rinke and W\"unsch \cite{rink81}.

Petrov type III space-times are characterized by the existence of
a spinor field $o^A$ satisfying \be \Psi_{ABCD}o^C
o^D=0\,,\;\;\;\; \Psi_{ABCD}o^D \neq 0. \ee Such a spinor field is
called a {\it repeated principal spinor} of the Weyl spinor and is
determined by the latter up to an arbitrary variable complex
factor. Let $\iota ^A$ be any spinor field satisfying \be o_A
\iota^A=1\,. \ee The ordered set ${o_A,\;\iota_A}$, called a dyad,
defines a basis for the 1-spinor fields on V${}_4$.

It was shown in \cite{car88} that the necessary conditions $III$
and $V$ imply that there exists a dyad $\{o_A, \iota_A\}$ and a
conformal transformation $\phi$ such that
\begin{eqnarray}
\label{4.100}
&&\kappa\,=\,\sigma\,=\,\rho\,=\,\tau\,=\,\epsilon\,=\,0\,,\\
\label{4.100b}
&&\Psi_0\,=\,\Psi_1\,=\,\Psi_2\,=\,\Psi_4\,=\,0\,,\;\Psi_3\,=\,-1\,,\\
\label{4.101}
&&\Phi_{00}\,=\,\Phi_{01}\,=\,\Phi_{02}\,=\,\Lambda\,=\,0\,,\\
\label{4.102}
&&\D \al\,=\,\D \beta\,=\,\D\pi\,=\,0\,,\\
\label{4.103} &&\de \Phi_{11}\,=\,\D \Phi_{11}\,=\,0\,.
\end{eqnarray}

We notice that the expressions (\ref{4.100b}) determine the tetrad
uniquely. On the other hand, conditions  (\ref{4.100}) are
invariant under any conformal transformation satisfying \be
\label{4.104} \D\phi\,=\,0\,,\qquad \delta \phi\,=\,0\,, \ee which
implies that we still have some conformal freedom. Under a
conformal transformation we have \cite{wal88}: \be \label{4.105}
\tilde{\Phi}_{11}=e^{-2\phi}\Phi_{11}\,. \ee Thus, we can choose
$\phi$ such that \be \label{4.106} \Phi_{11}\,=\,c\,, \ee where
$c$ is a constant. The conditions  (\ref{4.104}) are satisfied in
view of
 (\ref{4.103}).

Let us now derive some side relations that follow from the
previously obtained equations (\ref{4.100}) -- (\ref{4.106}) and
the necessary conditions (\ref{conIII}) -- (\ref{conVII}); these
will be required in the analysis of the following sections. We may
assume that $\al \beta\pi \neq 0$, since the case in which this is
not true was already considered in \cite{car88}.
By contracting condition $III$ with $\iota^{A}
 o^{B} \overline{\iota}^{\dot{A} \dot{B}}$
we get \be \label{4.107} \delta\beta=-\beta(\overline{\alpha} +
\beta)\,. \ee Using the Bianchi identities and the above
conditions, we obtain
\begin{eqnarray}
\label{4.108}
&&D\Phi_{12}=2 \overline{\pi} \Phi_{11}\,,\\
\label{4.109} &&D\Phi_{22}=-2(\beta+\overline{\beta})+2\Phi_{21}
\overline{\pi}+
2\Phi_{12} \pi\,,\\
\label{4.110}
&&\delta\Phi_{12}=2\overline{\alpha}+4\overline{\pi}+2\overline{\lambda}
\Phi_{11}
-2 \overline{\alpha}\Phi_{12}\,,\\
\label{4.111} &&\overline{\delta}\Phi_{12}=-2\beta+2\overline{\mu}
\Phi_{11}- 2\overline{\beta} \Phi_{12}\,.
\end{eqnarray}
The Ricci identities provide the following relevant Pfaffians:
\begin{eqnarray}
\label{4.112}
&&D\gamma=\opi \al +\beta \pi+\Phi_{11}\,,\\
\la{4.112b}
&&\D\lambda=(1/2)\delta \al-(11/2)\obe\al+\pi^2-2\pi\al-11\pi\obe-(3/2)\al^2\,,\\
\label{4.113}
&&\delta\opi=D\ola-\opi^2-\opi \oal+\opi\beta\,,\\
\label{4.114}
&&D\onu=\Delta\opi+\opi \omu +\ola \pi+\opi\oga-\opi\gamma-1+\Phi_{12}\,,\\
\label{4.115} &&\delta\alpha=\ode \beta+\alpha \overline{\alpha}+
\beta \overline{\beta}-2\beta \alpha + \Phi_{11}\,,\\
\label{4.116} &&\delta\pi=D\mu-\opi \pi +\pi \oal-\beta \pi\,.
\end{eqnarray}
We can obtain useful integrability conditions for the above
Pfaffians, by using Newman-Penrose (NP) commutation relations. By
substituting them in the commutator expression $[\delta,D]
\Phi_{22}-[\Delta,D] \Phi_{12}$, we get \be \label{4.117}
\delta\obe=-2\Phi_{11}-\obe \oal - 4\obe \opi-2D\omu-\beta \obe +2
\opi \pi.
 \ee
  By contracting condition $V$ with $ \iota^{ABCD}
\overline{\iota}^{\dot{A} \dot{B}} \overline{o}^{\dot{C}
\dot{D}}$, we find \be \label{4.118} 20\obe
\pi+12\obe\al+6\pi\al+3\al^2+\ode\al+2\ode\pi+\ode\obe+\obe^2=0\,.
\ee By substituting (\ref{4.106}) into this equation we get \be
\label{4.119}
\de(2\opi+\oal)=-20\opi\beta-11\beta\oal-6\opi\oal-3\oal^2\,. \ee
   From (\ref{4.115}), (\ref{4.116}) and (\ref{4.106}) we then
obtain: \be \label{4.120} \delta(2\pi +\alpha)=2\pi \oal +\alpha
\oal -6\beta \pi -3\beta\alpha-\Phi_{11}\,. \ee By contracting
condition $V$ with $ \iota^{ABC} o^D \overline{\iota}^ {\dot{A}
\dot{B} \dot{C} } \overline{o}^{\dot{D}}$, we find
\begin{eqnarray}
\nonumber && - 6\,\de\,\pi\, - 15\,\al\,\opi
 - 10\,\al\,\oal - 68\,\pi\,\opi - 15\,\pi\,\oal - 3
\,\de\,\al\, - 126\,\obe\,\beta \\
\nonumber
 & &  + 5\,\D\,\oga\, + 10\,\D\,\omu\, - 24\,\obe\,\oal - 3\,\ode\oal - 6\,
\ode\,\opi\, - 15\,\de\,\obe\, - 3\,\obe\,
\opi \\
\label{4.121a}
 & &  + 5\,\D\,{\gamma}\, + 10\,\D\,{\mu}\, - 15\,
\ode\beta - 24\,\beta\,{\al} - 3\,\beta\,\pi - 4\,\Phi_{11} =0\,.
\end{eqnarray}
Using (\ref{4.112}), (\ref{4.116}) and the complex conjugate of
(\ref{4.117}), we get \be \label{4.121} 9 \Phi_{11}+10 \obe \opi +
5[\D\mu+\D\omu] -2 \opi \al -12 \beta \obe -2 \al \oal -16 \opi
\pi +10 \beta \pi -2 \pi \oal =0\,. \ee On the other hand, the NP
commutator $ [\ode,\delta](\alpha+2\pi)=(\al-\obe)
\delta(\alpha+2\pi)+(-\oal + \beta)\ode(\alpha+2 \pi) $, yields
the following expression
\begin{eqnarray}
\nonumber &&2\pi \beta
\obe+22\pi\obe\oal+43\pi\obe\opi-22\opi\pi^2+\obe\D\mu+22\pi \D
\omu +
12\al\obe\oal\\
\label{4.122}
&&+6\obe\Phi_{11}-12\al\opi\pi+11\al\Phi_{11}+18\pi\Phi_{11}
+24\opi\obe\al+12\al\D\omu =0\,.
\end{eqnarray}
Eliminating $\D \omu$ between (\ref{4.121}) and (\ref{4.122}), and
solving for $\D \mu$, we get
\begin{eqnarray}
\nonumber &&\D
\mu=-\frac{1}{5}\left(108\,\pi\Phi_{11}-44\,\pi^{2}\oal-24\,\opi
\,{\al}^{2}-68\,\pi\al\oal\right.\\
\nonumber &&\-144\,\al\beta\obe+53\,\al\Phi_{11}
-274\,\pi\beta \obe+120\,\pi\beta\al-24\,{\al}^{2}\oal\\
\nonumber &&-242\,
\opi\,\pi^{2}+220\,\beta\pi^{2}-176\,\al\opi\,p-60\,\al
\obe\,\oal-30\,\obe\,\Phi_{11}\\
\label{4.123} &&\left.+5\,\pi\opi
\obe-110\,\oal\,\pi\obe\right)/(-\obe+12\,\al +22\,\pi)\,,
\end{eqnarray}
where we have assumed that the denominator of the expression
above, given by \be \la{den3.1} d_1:=-\obe+12\,\al+22\,\pi\,, \ee
is non-zero. The case $d_1=0$ will be considered later.

Substituting expression (\ref{4.123}) for $\D \mu$ into
(\ref{4.121}) we obtain
\begin{eqnarray}
\nonumber &&\D \omu =
-\frac{1}{5}\left(90\pi\Phi_{11}+12\beta\obe^2-10\obe^2\opi+
55\al\Phi_{11}+122\al\obe\opi-110\opi\pi^2-60\al\opi\pi\right.\\
\label{4.124} &&\left.+62\al\obe\oal+
21\obe\Phi_{11}+231\pi\obe\opi+112\oal\pi\obe\right)/
(-\obe+12\,\al+22\,\pi)\,.
\end{eqnarray}
One side relation can now be obtained by subtracting the complex
conjugate of
 (\ref{4.123}) from (\ref{4.124}). We obtain:
\begin{eqnarray}
 \nonumber
& &S_1 := {\displaystyle \frac {1}{5}}(720\,\oal^{ 2}\,\obe\,\al +
2904\,\pi^{2}\,\opi^{2} - 12\,\beta^{2}\, \obe^{2} +
288\,\oal^{2}\,\al^{2} + 528\,\pi^{2}\,\oal^{2} +
 528\,\opi^{2}\,\al^{2} \\ \nonumber
 & &  + 2420\,\oal\,\pi\,\obe\,\opi + 3056\,
\oal\,\pi\,\beta\,\obe + 2888\,\oal\,\al\,\opi\,\pi
 + 1320\,\oal\,\pi\,\beta\,\al + 1320\,\opi\,\beta\,\al^{2}\\
 \nonumber
 & &  + 1606\,\oal\,\beta\,\obe\,\al + 5802\,\opi\,\pi\,
\beta\,\obe + 1320\,\opi\,\oal\,\obe\,\al
 + 2420\,\opi\,\beta\,\al\,\pi + 1320\,\pi\,\obe
\,\oal^{2}\\
 \nonumber
 & &  + 3056\,\opi\,\beta\,\al\,\obe + 2552\,\pi\,
\opi^{2}\,\al + 305\,\beta\,\Phi_{11}\,\al + 570\,\beta\,\Phi_{11}
\,\pi - 51\,\beta\,\Phi_{11}\,\obe \\ \nonumber
 & &  + 24\,\oal\,\Phi_{11}\,\al - 86\,\oal\,\Phi_{11}\,
\pi + 305\,\oal\,\Phi_{11}\,\obe + 816\,\pi\,\al\, \oal^{2} +
2552\,\oal\,\opi\,\pi^{2} \\ \nonumber
 & &     \left.+ 816\,\oal\,\opi\,\al^{2} - 86\,\opi\,\al\,\Phi_{11}
- 396\,\opi\,\pi\,\Phi_{11} + 720\,\beta\,\oal
\,\al^{2} + 570\,\opi\,\obe\,\Phi_{11} \right/ \\
\label{4.124b} & &\qquad(\,(\, - 12\,\al - 22\,\pi +
\obe\,)\,(\,12\, \oal + 22\,\opi - \beta\,)\,)=0\,.
\end{eqnarray}

We notice that (\ref{4.123}) and (\ref{4.124})  have the same
denominator. Thus, if we keep these expressions for $\D \mu$ and
$\D \omu$,
 the Pfaffians $\delta\obe$,
$\delta \al$, $\delta \pi$, given by (\ref{4.117}), (\ref{4.113})
 and (\ref{4.116}), respectively, and
their complex conjugates, also have  the same denominator. This
procedure is crucial to keep the expressions to be obtained from
the integrability conditions within a reasonable size. Except for
$\ode \al$, all Pfaffians involving $\delta\,,\;\ode$, applied to
$\al\,,\;\beta\,,\pi$
 are explicitly determined.

The following expression for $\ode \al$ can be obtained from the
NP commutator $[\ode,\delta ]\obe=(\omu-\mu)\D \obe+
(\al-\obe)\delta\obe +(\beta-\oal)\ode \obe$ :
\begin{eqnarray}
\label{Ya1} \nonumber &&\ode \al =
(2\pi\D\omu-2\ode(\D\omu)-3\opi\al2-8\obe\D\omu-
11\al\obe\opi-2\opi\pi^2\\
\label{4.125}
&&-4\al\opi\pi-4\obe\Phi_{11}-14\opi\obe\opi)/\opi\,.
\end{eqnarray}
By substituting (\ref{4.123}) and (\ref{4.124}) into this equation
we get:
\begin{eqnarray}
\nonumber \lefteqn{{\ode \al} :=  - \,{\displaystyle \frac
{1}{5}}(19519\, \Phi_{11}\,\obe^{2}\,\al +
8570\,\Phi_{11}\,\obe\,\al^{2}
 + 1950\,\Phi_{11}\,\pi\,\al^{2} + 35850\,\Phi_{11}\,\pi\,\obe^{2}} \\
\nonumber
 & &  + 2900\,\Phi_{11}\,\pi^{2}\,\obe - 3180\,\pi^{2}
\,\obe^{2}\,\opi - 210\,\al^{3}\,\obe\,\opi + 150
\,\al^{3}\,\obe\,\oal - 1307\,\al^{2}\,\obe^{2}\,
\opi \\
\nonumber
 & &  - 180\,\al^{2}\,\beta\,\obe^{2} + 628\,\al^{2}\,
\obe^{2}\,\oal - 1280\,\pi^{2}\,\obe^{2}\,\oal +
1950\,\oal\,\obe^{3}\,\al + 4668\,\al\,\obe^{3}\,\beta
 \\
\nonumber
 & &  + 3520\,\obe^{3}\,\oal\,\pi + 8160\,\beta\,\pi
\,\obe^{3} + 330\,\obe^{3}\,\opi\,\pi + 975\,\Phi_{11}
\,\al^{3} - 860\,\obe^{3}\,\Phi_{11} \\
\nonumber
 & &  + 17950\,\Phi_{11}\,\obe\,\al\,\pi - 420\,\al^{2
}\,\obe\,\opi\,\pi + 300\,\al^{2}\,\obe\,\pi\,
\oal - 4116\,\al\,\obe^{2}\,\opi\,\pi \\
\nonumber
 & &  + 588\,\al\,\pi\,\oal\,\obe^{2} + 175\,
\obe^{3}\,\al\,\opi) \left/ {\vrule height0.37em width0em
depth0.37em} \right. \! \! (\obe^{2}\, \opi - 14\,\al\,\obe\,\opi
+ 10\,\obe\,\al\,
\oal + 20\,\obe\,\pi\,\oal \\
\label{4.126}
 & &  + 65\,\al\,\Phi_{11} - 12\,\beta\,\obe^{2} + 130\,
\pi\,\Phi_{11} - 31\,\obe\,\Phi_{11} - 28\,\obe\,\opi \,\pi -
6\,\obe^{2}\,\oal)\,,
\end{eqnarray}
where, for now, we assume that the denominator in the expression
above: \bea \non d_2&:=&\obe^{2}\, \opi - 14\,\al\,\obe\,\opi +
10\,\obe\,\al\,
\oal + 20\,\obe\,\pi\,\oal+ 65\,\al\,\Phi_{11} - 12\,\beta\,\obe^{2}\\
\la{den3.2} &&+ 130\,\pi\,\Phi_{11} - 31\,\obe\,\Phi_{11} -
28\,\obe\,\opi \,\pi - 6\,\obe^{2}\,\oal\,, \eea
 is non-zero.

Contracting condition $VII$ with
$\iota^{ABCDE}o^{F}\oo^{\dB\dC\dD}\oio^{\dA \dE \dF}$ yields
\begin{eqnarray}
\nonumber \lefteqn{{VII_{13}} := 648\,\obe^{2}\,\al\,\opi -
165\,\pi\, \opi\,\ode\obe - 40\,\pi\,\D\Phi_{21}
+ 1557\,\de\obe\,\al\,\pi} \\
\nonumber
 & &  + 270\,\al\,\obe\,\de\pi + 144\,
\de\al\,\al\,\obe + 378\,\al\,\obe\,\D
\oga + 2058\,\al^{2}\,\opi\,\pi \\
\nonumber
 & &  + 570\,\D\mu\,\ode\obe + 630
\,\D\mu\,\obe^{2} + 21\,\D\lambda\,\ode\oal - 201\,\D\oga\,\ode\pi \\
\nonumber
 & &  - 36\,\pi^{2}\,\ode\oal - 954\,\ode\beta\,\obe^{2} - 864\,\ode\beta\,\ode\obe -
864\,\beta\,\obe^{3} - 132\,\ode\al\,\ode\oal \\
\nonumber
 & &  + 1116\,\al^{3}\,\oal - 366\,\al^{2}\,\ode\oal
- 900\,\al\,\obe\,\de\obe - 20\,\al
\,\D\Phi_{21} + 1454\,\Phi_{11}\,\pi\,\obe \\
\nonumber
 & &  + 771\,\ode\opi\,\pi\,\obe + 423\,
\ode\al\,\oal\,\pi + 939\,\oal\,\al\,\ode
\pi - 1038\,\D\omu\,\al\,\pi \\
\nonumber
 & &  - 90\,\lambda\,\D(\,\D\omu\,) + 45\,
\pi\,\ode(\,\de\obe\,) - 15\,\pi\,\ode(\,
\D\oga\,) - 30\,\pi\,\ode(\,\D\omu\,) \\
\nonumber
 & &  + 84\,\al\,\obe\,\ode\opi - 774\,
\ode\al\,\obe\,\opi - 570\,\ode\obe
\,\oal\,\pi + 180\,\ode\al\,\beta\,\obe \\
\nonumber
 & &  + 135\,\D(\,\de\obe\,)\,\lambda - 45
\,\lambda\,\D(\,\D\oga\,) - 186\,\ode\,\obe\,\,\al\,\opi +
570\,\al\,\obe\,\D\omu
 \\
\nonumber
 & &  + 30\,\D\lambda\,\al\,\opi - 39\,
\D\lambda\,\oal\,\pi - 78\,\D\lambda\,\opi\,\pi -
744\,\oal\,\al\,\ode\obe - 924\,\oal\,\al^{2}\,\obe \\
\nonumber
 & &  - 387\,\pi^{2}\,\oal\,\obe - 21\,\D\lambda\,
\oal\,\al + 351\,\D\oga\,\pi\,\obe
 - 1773\,\obe\,\opi\,\ode\pi \\
\nonumber
 & &  + 216\,\oal\,\pi\,\ode\pi + 870\,\D\mu\,\al\,\obe - 198\,
\ode\oal\,\al\,
\obe + 762\,\ode\al\,\oal\,\al \\
\nonumber
 & &  + 378\,\beta\,\al\,\ode\obe + 828\,\al^{2
}\,\obe\,\beta - 1674\,\al\,\obe\,\ode\beta - 594\,\beta
\,\obe\,\ode\obe + 636\,\pi^{2}\,\oal\,\al
 \\
\nonumber
 & &  + 2289\,\al^{2}\,\pi\,\oal + 36\,\D\lambda\,\oal\,\obe -
660\,\al\,\pi\,\ode\oal + 1773\,\obe^{2}\,\opi\,\pi -
555\,\D\oga\,
\al\,\pi \\
\nonumber
 & &  + 2187\,\beta\,\pi\,\obe^{2} + 2619\,\beta\,\pi\,
\ode\obe - 888\,\oal\,\obe^{2}\,\al - 126\,
\ode\al\,\oal\,\obe \\
\nonumber
 & &  + 3681\,\beta\,\pi\,\al\,\obe - 2055\,\al\,\pi\,
\oal\,\obe - 6951\,\obe\,\opi\,\al\,\pi - 1476\,
\al\,\obe^{2}\,\beta \\
\nonumber
 & &  + 150\,\ode\al\,\opi\,\pi - 63\,\D
\lambda\,\obe\,\opi - 30\,\obe^{2}\,\oal\,\pi
 + 934\,\Phi_{11}\,\al\,\pi - 465\,\pi^{2}\,\obe\,\opi \\
\nonumber
 & &  - 1608\,\al\,\pi\,\ode\opi + 63\,\D
\lambda\,\beta\,\obe + 20\,\obe\,\D\Phi_{21}
 + 567\,\pi^{2}\,\beta\,\obe \\
\nonumber
 & &  - 426\,\ode\opi\,\ode\pi + 18
\,\D\lambda\,\D\omu - 39\,\pi^{2}\,\D\oga
+ 39\,\pi^{3}\,\oal - 78\,\pi^{2}\,\D\omu \\
\nonumber
 & &  + 78\,\pi^{3}\,\opi + 9\,\D\lambda\,\D
\oga + 324\,\de\al\,\ode\obe
 + 324\,\de\al\,\obe^{2} \\
\nonumber
 & &  - 189\,\ode\oal\,\ode\pi + 276
\,\ode\obe\,\D\gamma + 306\,\obe^{2}\,
\D\gamma + 432\,\al^{3}\,\opi \\
\nonumber
 & &  + 1080\,\al^{2}\,\de\obe - 144\,
\ode\al\,\ode\opi + 360\,\ode\al\,
\de\obe - 372\,\al^{2}\,\ode\opi \\
\nonumber
 & &  + 630\,\ode\obe\,\de\pi + 630
\,\obe^{2}\,\de\pi + 528\,\D\omu\,
\pi\,\obe + 408\,\opi\,\pi\,\ode\pi \\
\nonumber
 & &  + 246\,\al\,\opi\,\ode\pi + 639\,
\ode\oal\,\pi\,\obe + 156\,\al\,\obe\,\D\gamma
 - 2592\,\al^{2}\,\obe\,\opi \\
\nonumber
 & &  + 204\,\ode\al\,\al\,\opi + 117\,\beta\,
\obe\,\ode\pi - 657\,\de\obe\,\pi\,
\obe + 1128\,\pi^{2}\,\al\,\opi \\
\nonumber
 & &  - 108\,\ode\al\,\D\oga - 240
\,\ode\al\,\D\omu - 720\,\al^{2}\,\D
\omu - 324\,\al^{2}\,\D\oga \\
\nonumber
 & &  + 567\,\de\obe\,\ode\pi - 30
\,\D\lambda\,\ode\opi - 27\,\D\lambda
\,\de\obe + 117\,\pi^{2}\,\de\obe \\
\nonumber
 & &  - 378\,\D\omu\,\ode\pi - 102
\,\Phi_{11}\,\D\lambda + 182\,\Phi_{11}\,\pi^{2} + 366\,
\Phi_{11}\,\ode\pi \\
\nonumber
 & &  - 600\,\Phi_{11}\,\ode\obe - 600\,\Phi_{11}\,
\obe^{2} + 1008\,\Phi_{11}\,\al^{2} + 336\,\Phi_{11}\,
\ode\al + 1704\,\Phi_{11}\,\al\,\obe \\
\nonumber
 & &  + 30\,\pi\,\D(\,\ode\oga\,) - 90\,
\pi\,\de(\,\ode\obe\,) + 60\,\pi\,\D(\, \ode\omu\,) -
90\,\D(\,\ode\obe\,)
\,\mu \\
\nonumber
 & &  - 90\,\pi\,\D(\,\Delta\obe\,) - 660
\,\obe\,\pi\,\D\gamma - 270\,\al\,\ode(\,\de\obe\,)
 - 30\,\al\,\ode(\,\ode\oal\,) \\
\nonumber
 & &  - 60\,\al\,\ode(\,\ode\opi\,) + 30\,
\obe\,\ode(\,\D\gamma\,) + 60\,\obe\,\ode( \,\D\mu\,) +
90\,\al\,\ode(\,\D\oga
\,) \\
\nonumber
 & &  + 180\,\al\,\ode(\,\D\omu\,) - 180
\,\obe\,\ode(\,\de\pi\,) - 90\,\obe\,\ode (\,\de\al\,) +
45\,\pi\,\ode(\,\ode\oal
\,) \\
\nonumber
 & &  + 90\,\pi\,\ode(\,\ode\opi\,) + 30\,
\obe\,\D(\,\ode\gamma\,) + 60\,\obe\,\D( \,\ode\mu\,) +
15\,\lambda\,\D(\,\ode\oal
\,) \\
\nonumber
 & &  + 30\,\lambda\,\D(\,\ode\opi\,) + 60\,
\oal\,\ode(\,\ode\al\,) - 90\,\obe\,\ode(
\,\ode\beta\,) - 90\,\obe\,\D(\,\Delta\al\,) \\
\nonumber
 & &  - 180\,\obe\,\D(\,\Delta\pi\,) + 180
\,\beta\,\ode(\,\ode\obe\,) + 120\,\oal\,
\ode(\,\ode\pi\,) + 60\,\gamma\,\D(\,\ode\al\,) \\
\nonumber
 & &  + 120\,\gamma\,\D(\,\ode\pi\,) - 180
\,\oal\,\obe\,\ode\pi - 60\,\obe\,\D\mu\,\pi - 315\,\obe\,\ode\beta\,\pi \\
\label{4.127}
 & &  + 180\,\obe\,\de\pi\,\pi\,=\,0\,.
\end{eqnarray}
The second-order terms $\D(\ode\oga)$,  $\D(\ode\gamma)$, $\D(\ode
\omu)$, $\D(\Delta \obe)$, $\D(\Delta \al)$, $\D(\Delta \pi)$, can
be expressed in terms of known Pfaffians and $\ode \al$, by using
the NP commutation relations involving each pair of operators.
After the substitutions we obtain
\begin{eqnarray}
\nonumber \lefteqn{{VII_{13}} := {\displaystyle \frac
{2}{5}}(760\,\obe ^{2}\,\opi\,\pi\,\ode \al +
9600\,\obe\,\Phi_{11} \,\ode \al\,\pi + 25040\,\obe\,\ode \al\,
\oal\,\pi^{2}} \\
\nonumber
 & &  - 23440\,\obe\,\opi\,\pi^{2}\,\ode \al
+ 2400\,\beta\,\pi\,\ode \al\,\obe^{2} - 240\,
\obe^{2}\,\ode \al\,\oal\,\pi \\
\nonumber
 & &  + 2880\,\al\,\obe^{2}\,\beta\,\ode \al +
8600\,\al\,\Phi_{11}\,\ode \al\,\pi + 360\,\al\,\obe
^{2}\,\opi\,\ode \al \\
\nonumber
 & &  + 600\,\al\,\ode \al\,\oal\,\obe^{2}
 + 24880\,\al\,\obe\,\ode \al\,\oal\,\pi - 22160
\,\al\,\obe\,\opi\,\pi\,\ode \al \\
\nonumber
 & &  + 8940\,\al\,\Phi_{11}\,\ode \al\,\obe
 - 5280\,\al^{2}\,\obe\,\opi\,\ode \al + 6240\,
\obe\,\oal\,\al^{2}\,\ode \al - 9885300\,\al\,\Phi_{11}\,\pi^{2}\,\obe \\
\nonumber
 & &  + 113040\,\al\,\beta\,\obe^{4} - 1200\,\Phi_{11}\,
\ode \al\,\al^{2} + 48400\,\al\,\pi^{3}\,\Phi_{11} +
40200\,\Phi_{11}\,\pi\,\al^{3} \\
\nonumber
 & &  + 22400\,\Phi_{11}\,\ode \al\,\pi^{2} - 720
\,\obe^{3}\,\beta\,\ode \al - 300\,\ode \al\,
\oal\,\obe^{3} + 142635\,\al\,\obe^{3}\,\Phi_{11} \\
\nonumber
 & &  - 910512\,\al^{2}\,\obe^{3}\,\beta- 351912\,
\obe^{2}\,\al^{3}\,\opi - 5085800\,\obe\,\pi^{3}\,\Phi_{11}
 - 1467420\,\Phi_{11}\,\obe\,\al^{3} \\
\nonumber
 & &  - 1760\,\Phi_{11}\,\ode \al\,\obe^{2} -
3600\,\Phi_{11}\,\al^{4} + 35640\,\obe^{3}\,\al^{2}\,\opi
 - 360060\,\oal\,\obe^{3}\,\al^{2} \\
\nonumber
 & &  + 120000\,\Phi_{11}\,\pi^{2}\,\al^{2} - 633396\,
\Phi_{11}\,\obe^{2}\,\al^{2} + 8640\,\al^{3}\,\beta\,\obe^{2}
- 15840\,\al^{4}\,\obe\,\opi \\
\nonumber
 & &  - 155352\,\al^{3}\,\obe^{2}\,\oal + 18720\,
\al^{4}\,\obe\,\oal + 176880\,\pi^{2}\,\obe^{3}\,\opi
+ 205920\,\beta\,\pi\,\obe^{4} \\
\nonumber
 & &  + 85800\,\obe^{4}\,\oal\,\pi + 47100\,\oal\,
\obe^{4}\,\al + 227170\,\Phi_{11}\,\pi\,\obe^{3} -
932620\,\Phi_{11}\,\pi^{2}\,\obe^{2} \\
\nonumber
 & &  - 2702400\,\pi^{2}\,\beta\,\obe^{3} - 1041760\,\pi
^{3}\,\obe^{2}\,\oal - 2977440\,\pi^{3}\,\obe^{2}\,
\opi - 1064800\,\pi^{2}\,\obe^{3}\,\oal \\
\nonumber
 & &  - 835840\,\al^{2}\,\pi\,\oal\,\obe^{2} -
6584360\,\Phi_{11}\,\obe\,\al^{2}\,\pi + 74640\,\al^{3}\,
\obe\,\pi\,\oal \\
\nonumber
 & &  - 66480\,\al^{3}\,\obe\,\opi\,\pi + 75120\,
\pi^{2}\,\obe\,\al^{2}\,\oal - 70320\,\pi^{2}\,\al^{2}\,
\obe\,\opi - 2193544\,\al^{2}\,\obe^{2}\,\opi\,\pi \\
\nonumber
 & &  + 7200\,\al^{2}\,\beta\,\pi\,\obe^{2} - 4461264\,
\al\,\pi^{2}\,\obe^{2}\,\opi - 3128928\,\al\,\beta\,\pi\,
\obe^{3} \\
\nonumber
 & &  - 1237920\,\al\,\obe^{3}\,\oal\,\pi + 161720
\,\al\,\obe^{3}\,\opi\,\pi - 1662504\,\al\,\Phi_{11}\,\pi
\,\obe^{2}\\
\label{4.128} & &  - 1572528\,\al\,\pi^{2}\,\obe^{2}\,\oal-
25\,\obe^{4} \,\Phi_{11}) \left/
 \right. \! \! (\, - \obe + 12\,\al + 22\,\pi\,)^{2}\,=\,0\,.
\end{eqnarray}
Solving this equation for $\ode \al$ we get
\begin{eqnarray}
\nonumber \lefteqn{{\ode \al} :=  -
(227170\,\pi\,\obe^{3}\,\Phi_{11} + 40200\,\pi\,\al^{3}\,\Phi_{11}
+ 176880\,\pi^{2}\,\obe^{3}\,
\opi + 18720\,\obe\,\al^{4}\,\oal} \\
\nonumber
 & &  - 1467420\,\obe\,\al^{3}\,\Phi_{11} - 5085800\,
\obe\,\pi^{3}\,\Phi_{11} + 85800\,\pi\,\obe^{4}\,\oal
 - 1064800\,\pi^{2}\,\obe^{3}\,\oal \\
\nonumber
 & &  + 8640\,\obe^{2}\,\al^{3}\,\beta - 932620\,\pi^{2}
\,\obe^{2}\,\Phi_{11} - 15840\,\obe\,\al^{4}\,\opi +
205920\,\pi\,\obe^{4}\,\beta \\
\nonumber
 & &  - 2977440\,\pi^{3}\,\obe^{2}\,\opi - 1041760
\,\pi^{3}\,\oal\,\obe^{2} - 2702400\,\pi^{2}\,\beta\,
\obe^{3} - 360060\,\al^{2}\,\obe^{3}\,\oal \\
\nonumber
 & &  - 351912\,\al^{3}\,\obe^{2}\,\opi - 155352\,
\al^{3}\,\obe^{2}\,\oal - 633396\,\al^{2}\,\obe^{2}\,
\Phi_{11} - 910512\,\al^{2}\,\beta\,\obe^{3} \\
\nonumber
 & &  + 35640\,\al^{2}\,\obe^{3}\,\opi + 47100\,
\obe^{4}\,\al\,\oal + 142635\,\obe^{3}\,\al\,\Phi_{11}
 + 113040\,\obe^{4}\,\al\,\beta \\
\nonumber
 & &  - 3600\,\al^{4}\,\Phi_{11} - 1572528\,\pi^{2}\,
\obe^{2}\,\al\,\oal - 1662504\,\pi\,\obe^{2}\,\al\,
\Phi_{11} + 161720\,\pi\,\obe^{3}\,\al\,\opi \\
\nonumber
 & &  - 835840\,\pi\,\obe^{2}\,\al^{2}\,\oal -
1237920\,\pi\,\obe^{3}\,\al\,\oal - 4461264\,\pi^{2}\,\al
\,\obe^{2}\,\opi \\
\nonumber
 & &  - 2193544\,\al^{2}\,\pi\,\obe^{2}\,\opi +
7200\,\obe^{2}\,\al^{2}\,\pi\,\beta - 3128928\,\obe^{3}\,
\al\,\pi\,\beta \\
\nonumber
 & &  - 9885300\,\obe\,\pi^{2}\,\al\,\Phi_{11} + 75120
\,\obe\,\pi^{2}\,\al^{2}\,\oal - 70320\,\obe\,\pi^{2}
\,\opi\,\al^{2} \\
\nonumber
 & &  - 6584360\,\obe\,\al^{2}\,\pi\,\Phi_{11} + 74640
\,\obe\,\al^{3}\,\pi\,\oal - 66480\,\obe\,\al^{3}\,
\opi\,\pi + 48400\,\pi^{3}\,\al\,\Phi_{11} \\
\nonumber
 & &  + 120000\,\al^{2}\,\pi^{2}\,\Phi_{11} - 25\,\obe
^{4}\,\Phi_{11}) \left/ {\vrule height0.37em width0em depth0.37em}
\right. \! \! ( - 1760\,\obe^{2}\,\Phi_{11} - 300\,\obe
^{3}\,\oal - 720\,\beta\,\obe^{3} \\
\nonumber
 & &  + 360\,\obe^{2}\,\al\,\opi - 240\,\obe^{2}
\,\oal\,\pi + 2400\,\beta\,\pi\,\obe^{2} + 600\,\oal
\,\obe^{2}\,\al + 2880\,\al\,\obe^{2}\,\beta \\
\nonumber
 & &  + 760\,\obe^{2}\,\opi\,\pi - 5280\,\al^{2}\,
\obe\,\opi - 22160\,\obe\,\opi\,\al\,\pi + 25040
\,\pi^{2}\,\oal\,\obe \\
\nonumber
 & &  + 8940\,\obe\,\al\,\Phi_{11} + 9600\,
\obe\,\pi\,\Phi_{11} + 6240\,\oal\,\al^{2}\,\obe
- 23440\,\pi^{2}\,\obe\,\opi \\
\label{4.129}
 & &  + 24880\,\al\,\pi\,\oal\,\obe + 22400\,
\pi^{2}\,\Phi_{11} - 1200\,\al^{2}\,\Phi_{11} +
8600\,\pi\,\al\,\Phi_{11})\,,
\end{eqnarray}
where the denominator of (\ref{4.129}), \bea \non d_3 &:=& -
1760\,\obe^{2}\,\Phi_{11} - 300\,\obe
^{3}\,\oal - 720\,\beta\,\obe^{3} + 2880\,\al\,\obe^{2}\,\beta \\
\nonumber
 & &  + 360\,\obe^{2}\,\al\,\opi - 240\,\obe^{2}
\,\oal\,\pi + 2400\,\beta\,\pi\,\obe^{2} + 600\,\oal
\,\obe^{2}\,\al \\
\nonumber
 & &  + 760\,\obe^{2}\,\opi\,\pi - 5280\,\al^{2}\,
\obe\,\opi - 22160\,\obe\,\opi\,\al\,\pi + 25040
\,\pi^{2}\,\oal\,\obe \\
\nonumber
 & &  + 8940\,\obe\,\al\,\Phi_{11} + 9600\,
\obe\,\pi\,\Phi_{11} + 6240\,\oal\,\al^{2}\,\obe
- 23440\,\pi^{2}\,\obe\,\opi \\
\label{den3.3}
 & &  + 24880\,\al\,\pi\,\oal\,\obe + 22400\,
\pi^{2}\,\Phi_{11} - 1200\,\al^{2}\,\Phi_{11} +
8600\,\pi\,\al\,\Phi_{11}\,,
\end{eqnarray}
is assumed to be non-zero for now.

Subtracting (\ref{4.126}) from (\ref{4.129}) and taking the
numerator,
\begin{eqnarray}
\nonumber \lefteqn{{N_1} := 684288\,\obe^{5}\,\beta^{2}\,\al +
286000\,\pi^{3}\,\al\,\Phi_{11}^{2}} \\ \nonumber
 & &  - 7290900\,\pi^{2}\,\obe^{2}\,\Phi_{11}^{2} -
7913100\,\al^{3}\,\Phi_{11}^{2}\,\obe + 165600\,\obe^{5}
\,\oal^{2}\,\al \\ \nonumber
 & &  - 655680\,\pi^{3}\,\obe^{3}\,\oal^{2} +
295488\,\al^{3}\,\obe^{3}\,\opi^{2} - 205632\,\al^{3}\,
\obe^{3}\,\oal^{2} - 582090\,\pi\,\obe^{3}\,\Phi_{11} ^{2} \\
\nonumber
 & &  - 1157435\,\obe^{2}\,\Phi_{11}^{2}\,\al^{2} -
1517760\,\pi^{2}\,\obe^{4}\,\oal^{2} - 447840\,\obe^{
4}\,\oal^{2}\,\al^{2} \\ \nonumber
 & &  + 299000\,\pi^{2}\,\al^{2}\,\Phi_{11}^{2} + 123540\,
\obe^{5}\,\Phi_{11}\,\beta + 303600\,\pi\,\obe^{5}\,{\it ac }^{2}
+ 51450\,\obe^{5}\,\oal\,\Phi_{11} \\ \nonumber
 & &  - 126720\,\pi^{2}\,\obe^{4}\,\opi^{2} -
30643000\,\pi^{3}\,\obe\,\Phi_{11}^{2} - 23040\,\obe^{4}
\,\opi^{2}\,\al^{2} + 78000\,\pi\,\al^{3}\,\Phi_{11}^{2} \\
\nonumber
 & &  + 3111840\,\pi^{3}\,\obe^{3}\,\opi^{2} -
361718\,\obe^{3}\,\Phi_{11}^{2}\,\al + 25\,\obe^{5}\,
\opi\,\Phi_{11} + 1296000\,\pi\,\obe^{5}\,\beta^{2} \\ \nonumber
 & &  - 14398260\,\pi\,\obe^{2}\,\Phi_{11}\,\oal\,
\al^{2} - 57600\,\obe^{5}\,\oal\,\al\,\opi + 781920\,
\obe^{4}\,\oal\,\al^{2}\,\opi \\ \nonumber
 & &  + 89856\,\al^{3}\,\obe^{3}\,\oal\,\opi
 + 301945\,\obe^{4}\,\Phi_{11}^{2} - 138240\,\obe^{5}\,\beta\,\al\,\opi \\ \nonumber
 & &  + 1886976\,\obe^{4}\,\beta\,\al^{2}\,\opi -
1119744\,\obe^{4}\,\beta\,\oal\,\al^{2} + 682560\,\obe
^{5}\,\oal\,\al\,\beta \\ \nonumber
 & &  + 972820\,\obe^{4}\,\oal\,\al\,\Phi_{11} -
3755902\,\obe^{3}\,\oal\,\al^{2}\,\Phi_{11} - 266205\,
\obe^{4}\,\Phi_{11}\,\al\,\opi \\ \nonumber
 & &  + 1749108\,\obe^{4}\,\Phi_{11}\,\al\,\beta +
3627158\,\obe^{3}\,\Phi_{11}\,\al^{2}\,\opi - 3766320\,
\obe^{3}\,\beta\,\al^{2}\,\Phi_{11} \\ \nonumber
 & &  + 599880\,\al^{3}\,\Phi_{11}\,\obe^{2}\,\opi
 - 3021720\,\al^{3}\,\Phi_{11}\,\oal\,\obe^{2} + 6871680
\,\pi^{2}\,\obe^{4}\,\opi\,\beta \\ \nonumber
 & &  - 14919600\,\pi^{2}\,\obe^{3}\,\Phi_{11}\,\beta -
13552520\,\pi^{2}\,\obe^{3}\,\Phi_{11}\,\oal - 1389792\,
\pi^{2}\,\obe^{3}\,\oal^{2}\,\al \\ \nonumber
 & &  + 4357632\,\pi^{2}\,\obe^{3}\,\opi^{2}\,\al
 + 2915280\,\pi^{2}\,\obe^{4}\,\oal\,\opi + 13578100
\,\pi^{2}\,\obe^{3}\,\Phi_{11}\,\opi \\ \nonumber
 & &  - 3718080\,\pi^{2}\,\obe^{4}\,\oal\,\beta -
60662200\,\pi^{2}\,\obe\,\Phi_{11}^{2}\,\al - 713952\,\pi^{2}
\,\obe^{3}\,\opi\,\al\,\oal \\ \nonumber
 & &  - 8470120\,\pi^{2}\,\obe^{2}\,\Phi_{11}\,\opi
\,\al - 22343840\,\pi^{2}\,\obe^{2}\,\Phi_{11}\,\al\, \oal +
46000\,\pi^{2}\,\al^{2}\,\Phi_{11}\,\oal\,\obe \\ \nonumber
 & &  - 64400\,\pi^{2}\,\al^{2}\,\Phi_{11}\,\obe\,
\opi - 26400\,\pi^{2}\,\al\,\Phi_{11}\,\beta\,\obe^{2} -
105600\,\pi\,\obe^{5}\,\oal\,\opi \\ \nonumber
 & &  + 1806460\,\pi\,\obe^{4}\,\Phi_{11}\,\oal -
38468250\,\pi\,\obe\,\Phi_{11}^{2}\,\al^{2} - 111360\,\pi\,
\obe^{4}\,\opi^{2}\,\al \\ \nonumber
 & &  + 3379920\,\pi\,\obe^{4}\,\Phi_{11}\,\beta -
474200\,\pi\,\obe^{4}\,\Phi_{11}\,\opi - 253440\,\pi\,
\obe^{5}\,\opi\,\beta \\ \nonumber
 & &  + 1268640\,\pi\,\obe^{5}\,\oal\,\beta -
1647600\,\pi\,\obe^{4}\,\oal^{2}\,\al - 6010420\,\pi\,
\obe^{2}\,\Phi_{11}^{2}\,\al \\ \nonumber
 & &  - 939744\,\pi\,\obe^{3}\,\oal^{2}\,\al^{2}
 + 1993248\,\pi\,\obe^{3}\,\opi^{2}\,\al^{2} + 3021840\,
\pi\,\obe^{4}\,\oal\,\opi\,\al \\ \nonumber
 & &  + 51264\,\pi\,\obe^{3}\,\opi\,\oal\,\al
^{2} + 14037554\,\pi\,\obe^{3}\,\Phi_{11}\,\al\,\opi -
14255872\,\pi\,\obe^{3}\,\Phi_{11}\,\oal\,\al \\ \nonumber
 & &  - 15000840\,\pi\,\obe^{3}\,\Phi_{11}\,\al\,\beta
 - 592500\,\pi\,\obe^{2}\,\Phi_{11}\,\al^{2}\,\opi -
4074624\,\pi\,\obe^{4}\,\oal\,\al\,\beta \\ \nonumber
 & &  + 7198848\,\pi\,\obe^{4}\,\opi\,\al\,\beta -
14400\,\pi\,\al^{2}\,\Phi_{11}\,\beta\,\obe^{2} - 16800\,
\pi\,\al^{3}\,\Phi_{11}\,\obe\,\opi \\ \nonumber
 & &  + 12000\,\pi\,\al^{3}\,\Phi_{11}\,\oal\,\obe
 - 9855600\,\pi^{3}\,\obe^{2}\,\Phi_{11}\,\opi - 11178800
\,\pi^{3}\,\obe^{2}\,\Phi_{11}\,\oal \\
\label{4.130}
 & &  - 929760\,\pi^{3}\,\obe^{3}\,\opi\,\oal
 + 44000\,\pi^{3}\,\al\,\Phi_{11}\,\oal\,\obe - 61600\,
\pi^{3}\,\al\,\Phi_{11}\,\obe\,\opi=0\,.
\end{eqnarray}

\section{The case $\Phi_{11}=0$}
\la{sec-3}
 Carminati and
McLenaghan \cite{car88} used the  conditions III and V given in
Section \ref{sec-2} to prove that Huygens' principle is not
satisfied if any of the spin coefficients $\al$, $\beta$ or $\pi$
vanish. We now  extend the proof for the case in which
$\al\beta\pi \neq 0$
 and $\Phi_{11}=0$; i.e., we shall prove the following
theorem:
\begin{theorem}
\la{teo3} Let V${}_4$ be any space-time which admits a spinor dyad
with the properties \be o_{A;B \dot{B}}=o_AI_{B \dot{B}}\,, \ee
where $ I_{B \dot{B}} $ is a 2-spinor, and \be \Psi_{ABCD;E
\dot{E}}\,\iota^A \iota^B \iota^C o^D o^E \overline{o}^{\dot{E}
}=0\,, \ee \be R=0\,,\qquad\Phi_{AB \dot{A} \dot{B}}o^A o^B=0\,.
\ee Then the validity of Huygens' principle for the conformally
invariant equation (\ref{selfad}) implies that \be \la{eq1teo3}
\Phi_{AB\dA\dB}o^A\iota^B\oo^{\dA}\oio^{\dB}\neq 0\,. \ee
\end{theorem}
\noindent
{\bf Proof:}\\
When  $\Phi_{11}=0$ the quantity $N_1$, given by (\ref{4.130}),
factors in the following form : \be \la{eq-N1}
N_1:=-12\obe\,p_1\,p_2\,, \ee where \be \la{eq-p1}
p_1:=12\,\beta\,\obe + 2\,\pi\,\oal + 2\,\al\,\oal
 + 5\,\obe\,\oal + 6\,\opi\,\pi + 2\,\opi\,\al\,,
\ee
\begin{eqnarray}
\nonumber p_2&:=& 1188\,\obe\,\beta\,\al+ 240\,\al\,\obe\,\opi +
440 \,\obe\,\opi\,\pi- 1265\,\obe\,\pi\,\oal\\ \nonumber
 & &  - 2250\,\obe\,\beta\,\pi + 6830\,\pi^{2}\,\oal
 + 7647\,\al\,\pi\,\oal + 2142\,\al^{2}\,\oal  \\
\la{eq-p2}
 & &  - 690\,\obe\,\al\,\oal - 10805\,\opi
\,\pi^{2} - 11529\,\al\,\opi\,\pi - 3078\,\opi\,\al^{2}\,.
\end{eqnarray}

Let us consider first the case in which $p_2=0$. Applying $\ode$
to (\ref{eq-p2}) and solving for $\ode \al$, we obtain:
\begin{eqnarray}
\nonumber \lefteqn{{\ode \al} :=  - (690120\,\obe\,\beta\,\al^{3}
- 177100 \,\obe^{3}\,\oal\,\pi + 2475000\,\obe\,\al\,\pi^{2}\,
\beta - 97175\,\oal\,\obe^{3}\,\al} \\
\nonumber
 & &  - 186390\,\al\,\obe^{3}\,\beta + 1716210\,\oal
\,\al^{2}\,\obe^{2} - 1131915\,\al^{2}\,\obe^{2}\,
\opi - 4470219\,\obe\,\al^{3}\,\oal \\
\nonumber
 & &  - 3875990\,\pi^{2}\,\obe^{2}\,\opi - 341280
\,\beta\,\pi\,\obe^{3} + 5791820\,\pi^{2}\,\oal\,
\obe^{2} + 9784170\,\al^{3}\,\opi\,\pi \\
\nonumber
 & &  + 2573586\,\al^{2}\,\beta\,\obe^{2} + 8639361\,\al
^{3}\,\obe\,\opi + 8903160\,\pi^{2}\,\beta\,\obe^{2} -
28683640\,\pi^{3}\,\obe\,\oal \\
\nonumber
 & &  - 11970480\,\pi^{2}\,\al^{2}\,\oal + 18687060\,{
p}^{2}\,\opi\,\al^{2} - 6346350\,\al^{3}\,\pi\,\oal +
9567000\,\beta\,\pi\,\al\,\obe^{2} \\
\nonumber
 & &  + 2615220\,\obe\,\beta\,\pi\,\al^{2} + 33800\,
\obe^{3}\,\al\,\opi + 61600\,\obe^{3}\,\opi\,\pi
 + 90203190\,\pi^{2}\,\al\,\obe\,\opi \\
\nonumber
 & &  - 1119420\,\al^{4}\,\oal - 4188240\,\obe^{2}
\,\opi\,\al\,\pi + 6302070\,\al\,\obe^{2}\,\pi\,\oal
 \\
\nonumber
 & &  - 24866544\,\obe\,\al^{2}\,\pi\,\oal -
46210320\,\obe\,\al\,\pi^{2}\,\oal + 48333948\,\al^{2}\,
\obe\,\opi\,\pi \\
\nonumber
 & &  + 56154880\,\pi^{3}\,\obe\,\opi +
1705860\,\al^{4}\,\opi - 7513000\,\pi^{3}\,\al\,\oal + 11885500\,
\opi\,\pi^{3}\,\al) \left/ {\vrule
height0.37em width0em depth0.37em} \right. \! \!  \\
\non
 & &  \left((-12\al-22\pi+bc)(115\obe\oal+126\beta\obe -40\obe\opi
+783\opi\al-1634\pi\oal \right.\\
\la{4.133} &&\left. +1448\pi\opi-921\al\oal)\right)\,,
\end{eqnarray}
where the denominator of the expression above, given by \bea
\nonumber d_4&:=&(-12\al-22\pi+\obe)(115\obe\oal+126\beta\obe
-40\obe\opi
+783\opi\al-1634\pi\oal \\
\label{den3.4} && +1448\pi\opi-921\al\oal)\,, \eea is assumed to
be non-zero, for now.

Here $N_1$, and  all equations obtained by comparing different
expressions for $\ode \oal$, are polynomials in three complex
variables $\al$, $\beta$ and $\pi$. One complex variable can be
eliminated by  introducing the following new variables: \be
\la{x_1} x_1:=\frac{\al}{\pi}\,,\qquad x_2:=\frac{\beta}{\opi}\,.
\ee In what follows we first prove that the necessary conditions
imply that both $\x$ and $\y$ are constants. Then, later, we shall
prove that this leads to a contradiction.

In the new variables defined by (\ref{x_1}), the expression
(\ref{eq-p2}) assumes the form
\begin{eqnarray}
\nonumber \lefteqn{{p_2}= - 2250\,\yc\,\y - 1188\, \yc\,\y\,\x -
3078\,\x^{2} - 11529\,\x
 - 10805 } \\
\nonumber
 & &  + 7647\,\x\,\xc + 6830\,+ 2142\,\x^{2}\,\xc\xc - 1265
\,\yc\,\xc + 440\,\yc - 690\,\yc\,\x\,\xc \\
\label{4.134}
 & &  + 240\,\x\,\yc+ 2142\,\x^{2}\,\xc=0\,.
\end{eqnarray}
Subtracting (\ref{4.133}) from (\ref{4.129}) (with $\Phi_{11}=0$),
and taking the numerator, gives
\begin{eqnarray}
\nonumber \lefteqn{{N_2} :=  - 75130000\,\xc^{2}\,\x -
568034312\,\yc\,\x\,\xc^{2} - 162662000\,\x^{2}\,\xc^{2}} \\
\nonumber
 & &  - 263829680\,\yc\,\xc^{2} + 69828000\,
\xc\,\yc\,\y\,\x + 91299060\,\yc^{2} \\ \nonumber
 & &  - 105963000\,\yc\,\y\,\x + 328900
\,\yc^{4}\,\xc^{2} - 3248115\,\yc^{3} - 97977600\,\x^{4} \\
\nonumber
 & &  - 12927600\,\x^{5} - 278399100\,\x^{3}
 + 37400\,\yc^{4} - 16070400\,\x^{4}\,\yc\,
\y \\ \nonumber
 & &  - 277285752\,\x^{2}\,\yc^{2}\,\y
 + 6646212\,\x^{2}\,\yc^{3}\,\y + 78647544\,
\x^{4}\,\yc\,\xc \\ \nonumber
 & &  - 48437136\,\x^{3}\,\yc^{2}\,\y +
15461670\,\x\,\xc\,\yc^{3} - 39517398\,\x ^{3}\,\yc^{2}\,\xc \\
\nonumber
 & &  + 12081744\,\xc\,\x^{3}\,\yc^{2}\,
\y + 593329572\,\x^{3}\,\yc\,\xc + 4160700\,\x^{2}\,\yc^{3}\,\xc
\\ \nonumber
 & &  - 221403987\,\x^{2}\,\yc^{2}\,\xc
 - 42924720\,\xc\,\yc^{3}\,\y - 459117172\,
\xc^{2}\,\yc\,\x^{2} \\ \nonumber
 & &  + 1679968716\,\xc\,\yc\,\x^{2} +
112691520\,\xc\,\yc^{2}\,\y \\ \nonumber
 & &  + 163297776\,\xc\,\yc^{2}\,\y\,
\x - 413617894\,\xc\,\yc^{2}\,\x + 125141836\,\xc^{2}\,\yc^{2}\,\x
\\ \nonumber
 & &  + 14366310\,\xc\,\yc^{3} - 169509600\,
\x^{2}\,\yc\,\y - 15940800\,\x^{2}\,\yc^{2}\,\y^{2} \\
\nonumber
 & &  + 114001200\,\x^{2}\,\yc\,\y\,
\xc + 62078400\,\x^{3}\,\yc\,\y\,\xc \\ \nonumber
 & &  + 11275200\,\x^{4}\,\yc\,\y\,\xc
 - 45650088\,\x\,\xc\,\yc^{3}\,\y -
213660\,\yc^{4}\,\x\,\y \\ \nonumber
 & &  + 77697000\,\xc\,\x^{2}\,\yc^{2}\,
\y - 12134448\,\xc\,\x^{2}\,\yc^{3}\, \y \\ \nonumber
 & &  - 165106878\,\xc^{2}\,\x^{3}\,\yc
 + 66914052\,\xc^{2}\,\x^{2}\,\yc^{2} +
383646000\,\x^{3}\,\xc \\ \nonumber
 & &  + 136587600\,\x^{4}\,\xc - 1405233644\,
\x^{2}\,\yc - 495836310\,\x^{3}\,\yc \\ \nonumber
 & &  + 78021681\,\x^{2}\,\yc^{2} - 257647540
\,\xc\,\yc^{2} - 44763192\,\y^{2}\,\yc^{3}\,\x \\ \nonumber
 & &  - 14850000\,\y^{2}\,\x\,\yc^{2} -
121050\,\yc^{4}\,\x\,\xc - 11901168\,\x^{2}\,\yc^{3}\,\y^{2} \\
\nonumber
 & &  - 4276800\,\x^{3}\,\yc^{2}\,\y^{2}
 - 22290588\,\xc^{2}\,\x^{4}\,\yc + 11929896\,
\xc^{2}\,\x^{3}\,\yc^{2} \\ \nonumber
 & &  - 3427140\,\xc^{2}\,\x^{2}\,\yc^{3}
 - 65630268\,\x^{4}\,\yc + 13886406\,\x^{3}\,
\yc^{2} \\ \nonumber
 & &  - 939210\,\x^{2}\,\yc^{3} - 11744220\,
\xc^{2}\,\yc^{3} + 78035680\,\xc^{2}\,\yc ^{2} +
20400\,\yc^{4}\,\x \\ \nonumber
 & &  + 1533600\,\y^{2}\,\yc^{4} + 146164809\,
\yc^{2}\,\x + 750960\,\x\,\yc^{4}\,\y\,\xc \\ \nonumber
 & &  + 179400\,\xc^{2}\,\yc^{4}\,\x -
90396000\,\x^{3}\,\yc\,\y - 12687610\,\x \,\xc^{2}\,\yc^{3} \\
\nonumber
 & &  + 25076106\,\yc^{3}\,\x\,\y -
528822552\,\yc^{2}\,\x\,\y - 3492935\,\yc ^{3}\,\x \\ \nonumber
 & &  + 23644920\,\yc^{3}\,\y -
397830\,\yc^{4}\,\y - 336008520\,\yc^{2}\,\y - 6426000
\,\xc^{2}\,\x^{5} \\ \nonumber
 & &  - 131992500\,\x^{3}\,\xc^{2} + 18230400
\,\x^{5}\,\xc + 824256\,\x\,\yc^{4}\, \y^{2} \\ \nonumber
 & &  + 1395480\,\yc^{4}\,\xc\,\y -
221925\,\yc^{4}\,\xc - 47574000\,\x^{4}\, \xc^{2} - 166397000\,\x
\\ \nonumber
 & &  - 42109200\,\y^{2}\,\yc^{3} - 836819360
\,\yc - 1770562898\,\x\,\yc + 224037000\,\x\,\xc \\
\nonumber
 & &  + 478803300\,\x^{2}\,\xc - 351507100\,
\x^{2} + 999889240\,\yc\,\xc \\
\label{4.135}
 & &  + 2115717928\,\yc\,\x\,\xc =0\,.
\end{eqnarray}

We now wish to determine the solutions of the system of algebraic
equations $\{p_2=0,\, N_2=0\}$. This may be accomplished in
principle using the Gr\"{o}bner basis method of Buchberger
\cite{gedd92} as follows.  First, we treat the quantities $\x$,
$\y$, $\xc$, $\yc$ as independent variables, as view the
quantities $p_2$, $N_2$ as polynomials in these indeterminates
over the field of rational numbers.  (In the subsequent analysis
we may use the fact that some variables are complex conjugates of
each other, but this will not be necessary for our immediate
purpose.) Then, by computing a Gr\"{o}bner basis for the set
$\{p_2,\, N_2\}$ (actually, the ideal $<p_2,\, N_2>$) with respect
to a purely lexicographic ordering of terms (see \cite{gedd92}) we
obtain a new set of polynomials with the same solutions but in
which the variables have been successively eliminated as far as
possible.  In order to speed the computations, we use a special
variant of the algorithm \cite{cza89} which combines the nonlinear
elimination with factorization of intermediate results. (This
algorithm is available in the Maple system as the function {\tt
gsolve}.) For the polynomials $\{p_2,\, N_2\}$, the algorithm
produces the following components, which collectively contain all
solutions: \be \label{4.137} G_{1} := [\, - 8 + 23\,\xc, 11\,\yc +
8, 66\, \x + 125\,]\,, \ee \be \label{4.138} G_{2} :=
[\,9108\,\y\,\yc + 247,  - 8 + 23\,\xc, 66\,\x + 125\,]\,, \ee \be
\label{4.139} G_{3} := [\,828\,\y\,\yc + 75\,\yc + 77,
 - 8 + 23\,\xc, 66\,\x + 125\,]\,,
\ee \be \label{4.140} G_{4} := [\,271\,\x + 138\,\y\,\yc + 517,
 - 8 + 23\,\xc\,]\,,
\ee \be \label{4.141} G_{5} := [\,36\,\y\,\yc + 7 - 5\,\xc, 6\, \x
+ 11\,]\,, \ee
\begin{eqnarray}
\nonumber \lefteqn{G_{6} := [671514624\,\y^{2}\,\yc^{2} +
488374272\,\y^{2}\,\yc - 220446720\,\xc\,\y} \\ \nonumber
 & &  + 35785728\,\yc^{2}\,\y + 88473600\,
\y\,\yc - 27979776\,\x\,\y + 69101568\, \y \\ \nonumber
 & &  - 167878656\,\yc^{2}\,\xc^{2} +
181020672\,\xc^{2}\,\yc - 26599040\,\xc^{2} \\ \nonumber
 & &  + 73852416\,\yc^{2}\,\xc - 132857600\,
\xc\,\yc + 23168456\,\xc - 26978094\,\x
 \\ \nonumber
 & &  - 49722705 - 2204136\,\x^{2} + 48043776\,
\x\,\yc + 111324800\,\yc - 7645440\,\yc^{2},
 \\ \nonumber
 & & 847872\,\xc\,\yc\,\y - 294912\,\y\,
\yc + 423936\,\xc^{2}\,\yc - 382720\,\xc ^{2} \\ \nonumber
 & &  - 218112\,\xc\,\yc + 323928\,\xc
 - 54450\,\x - 169493 + 24576\,\yc,  \\ \nonumber
 & & 139392\,\yc^{3}\,\y + 202752\,\yc^{2}\,
\y + 73728\,\y\,\yc + 69696\,\yc^{3}\, \xc \\ \nonumber
 & &  + 38456\,\yc^{2}\,\xc - 54656\,\xc
\,\yc - 33280\,\xc - 4224\,\x + 10432 - 11616\, \x\,\yc \\
\nonumber
 & &  + 22544\,\yc + 2827\,\yc^{2} - 11616\,
\yc^{3} - 7986\,\yc^{2}\,\x, 304128\,\x\, \y\,\yc \\ \nonumber
 & &  + 576000\,\y\,\yc + 66240\,\xc\,
\yc - 59800\,\xc - 191598\,\x - 279575 - 17424 \,\x^{2} \\
\label{4.142}
 & &  + 198000\,\x\,\yc + 351960\,\yc,
192\,\x\,\xc - 282\,\x - 505 + 280\,\xc]\,.
\end{eqnarray}
Using the fact that the pairs $(\x,\,\y)$ and $(\xc\,,\,\yc)$ are
complex conjugates of each other, we conclude that the sets $G_1$
to $G_5$ provide solutions which are either impossible or in which
$x_1$ and $x_2$ are constant. In the case of $G_6$, this is not
immediately obvious. Its smallest term is: \be \label{4.143}
192\,\x\,\xc - 282\,\x - 505 + 280\,\xc =0\,. \ee Subtracting
(\ref{4.143}) from its complex conjugate we obtain the conclusion
that $x_1$ is real, which implies that it must be constant. It
follows that $x_2$ must be constant as well.

Let us consider now the case \be \label{4.144}
 p_1=12\y\yc+6+2\x+2\xc+2\x\xc+5\xc\yc=0\,.
\ee We then use the side relation $S_1$ given by (\ref{4.124b}),
whose numerator takes the form:
\begin{eqnarray}
\nonumber &&p_{3} :=  - 6\,\y^{2}\,\yc^{2} + 1210\,\y\, \x +
1276\,\x + 1276\,\xc + 360\,\y\,
\x^{2}\,\xc  \\
\nonumber &&+ 2901\,\y\,\yc + 1528\,\x\,\y \,\yc +
660\,\x\,\yc\,\xc + 660\,\y
\,\x^{2} + 408\,\x\,\xc^{2}\\
\nonumber &&  + 660\,\xc^{2}\,\yc + 408\, \xc\,\x^{2} +
264\,\xc^{2} + 144\,\x^{2}
\,\xc^{2} + 1452 + 264\,\x^{2} \\
\nonumber
&&+ 1210\,\xc\,\yc + 1444\,\xc\,\x + 1528\,\y\,\xc\,\yc + 660\,\y\,\xc\,\x\\
\label{4.146} &&+ 803\,\y\,\x\,\yc\, \xc+ 360\,\x\, \yc\,\xc^{2} =
0\,.
\end{eqnarray}
Applying our nonlinear elimination algorithm as before to {$p_1$ ,
$p_3$} we obtain the following equivalent system of equations:
\begin{eqnarray}
\label{4.148}
&&s_1:=6\y\yc+31\x\xc+56\x+\xc+3\,=0,\\
\label{4.149}
&&s_2:=72\y\x\xc+132\x\y+31\x\xc^2+56\xc\x+\xc^2+3\xc=0\,,\\
\label{4.150} &&s_3:=\xc\yc-22\x-12\xc\x=0\,.
\end{eqnarray}
Subtracting (\ref{4.148}) from its complex conjugate yields
$\xc=\x$.
 Subtracting (\ref{4.150}) from its
complex conjugate now gives $\y=\yc=12\x+22$. Substituting these
relations back in (\ref{4.148}) and (\ref{4.149}) results in a
system with no solution.

It thus follows that in either of the cases which arise from
equation (\ref{eq-N1}), $x_1$ and $x_2$ must necessarily be
constant. However, it may be shown (though we postpone the details
until the following section) that this too leads to a
contradiction.

We must finally consider the case in which the denominator of
$D\mu$, given by (\ref{4.123}), is zero. Here we shall suppose
that $\Phi_{11}$ is not necessarily zero, so that the side
relations derived in Section \ref{sec-2} will remain valid in the
following section as well. According to (\ref{den3.1}) and
(\ref{x_1}), \be \label{4.195} d_1:=22+12\x-\yc=0\,. \ee From
(\ref{4.123}) we obtain \bea \nonumber && E_1 :=  -
53\,\x\,\phi_{11} + 242 - 220\, \y + 24\,\xc\,\x^{2} +
30\,\yc\,\phi_{11}
 + 274\,\y\,\yc \\
\nonumber
 & & \mbox{} - 5\,\yc + 110\,\yc\,\xc + 24\,
\x^{2} + 44\,\xc + 144\,\x\,\y\,\yc
 - 108\,\phi_{11}\\
\label{4.196}
 & & - 120\,\x \y + 68\,\x\,\xc + 176\,\x + 60\,\xc\,\yc\,\x=0\,,
\eea where $\phi_{11}$ is defined as follows: \be \label{4.197}
\phi_{11}:=\frac{\Phi_{11}}{\pi \opi}\,. \ee Applying $\ode$ to
$f_1$, using (\ref{4.106}), (\ref{4.113}) and (\ref{4.112}), and
solving for $\ode \al$, we get \be \label{4.198} \ode \al=
120\obe\al+66\pi\al+220\pi\obe+33\al^2-\obe^2\,. \ee By applying
$\de$ to  $d_1$, now using (\ref{4.115}), (\ref{4.116}),
(\ref{4.117}) and (\ref{4.121}) and solving for $\D\mu$,  we get
\be \label{4.199} \D\mu=(4\opi\al+29\beta\obe
+64\al\oal-370\beta\pi+32\opi\pi-190\beta\al+
114\pi\oal+5\obe\oal-68\Phi_{11})/20\,. \ee Subtracting $D\omu$,
given by  (\ref{4.121}), from the complex conjugate of
(\ref{4.199}), gives
\begin{eqnarray}
\nonumber E_2 &:=& 20\,\phi_{11} - 2\,\y\,\yc - 22 \,\x -
24\,\x\,\xc + 66\,\y +35\,\x\,\y - 22\,
\xc\\
\label{4.200}
 & &  + 35\,\yc\,\xc + 66\,\yc=0\,.
\end{eqnarray}
Applying $\ode$ to (\ref{4.200}) gives
\begin{eqnarray}
\nonumber E_3 &:=& 1168\,\x\,\phi_{11} + 3980\,\y\, \x^{2} - 264 +
14520\,\y - 3584\,\xc\,\x
^{2} - 255\,\yc\,\phi_{11} \\
\nonumber &&  - 2838\,\y\,\yc - 9482\,\x\,\xc - 24200\,\yc -
12870\,\yc\,\xc - 12870\,
\x\,\yc \\
\nonumber
 & &  - 3784\,\x^{2} - 1210\,\yc^{2} - 4928\,
\xc - 1514\,\x\,\y\,\yc + 20\,\y\,
\yc^{2} - 625\,\yc^{2}\,\xc \\
\label{4.201}
 & &  + 2046\,\phi_{11} + 15180\,\x\,\y - 7480
\,\x - 6835\,\xc\,\yc\,\x=0\,.
\end{eqnarray}
Applying nonlinear elimination  to $d_1$, $\overline{f_1}$, $E_1$,
$\overline{E_1}$,
 $E_2$ and $E_3$, we find that this system has no solution.

The cases where each of the denominators $d_2$, $d_3$ and $d_4$,
that appeared in the preceding equations are zero lead to
contradictions, according to \cite{sas97}. The demonstration of
this fact follows the steps described above and will not be
presented here for brevity.

Thus, for Huygens' principle to be satisfied on Petrov type III
space-times we must have $\Phi_{11}\neq 0$, and Theorem
\ref{teo3}, which states this result in a conformally invariant
way, is proved.


\section{The case $\Phi_{11} \neq 0$}
\la{sec-4} We shall now examine the sole case which remains after
the analysis of the previous section, namely that in which
$\alpha\beta\pi \neq 0$ and $\Phi_{11} \neq 0$.  This, in view of
Theorem \ref{teo3}, will complete the proof of Theorem
\ref{teo-main}. Our approach is related to that of the previous
section, in that we reduce the problem to an issue of solvability
of a purely algebraic system of equations.

We first observe that, in addition to the algebraic equations
given by (\ref{4.124b}) and (\ref{4.130}), an extra independent
equation may be obtained by applying the NP operator $\delta$ to
(\ref{4.124b}). All of the Pfaffians which result are known
explicitly, and may be replaced using the expressions found in
Section \ref{sec-2} to obtain a (very large) expression in the
complex variables $\alpha$, $\beta$, and $\pi$ and the real
quantity $\Phi_{11}$. Upon transforming variables according to
(\ref{x_1}) and (\ref{4.197}), we obtain a complex quantity in the
new variables $\x$, $\y$, $\xc$, $\yc$, and $\phi_{11}$. (This
polynomial contains 408 terms of maximum total degree 9.) Together
with the equations which similarly follow from (\ref{4.124b}) and
(\ref{4.130}), we have in effect a system of five equations in
five real variables. Let us denote the set of polynomials which
arise in this system (i.e. when the equations are written with a
right hand side of $0$) by $F$.

It must be mentioned that the approach of the previous section,
namely computing the solutions by explicit elimination, is
impossible in the present case due to the intrinsic computational
complexity of nonlinear elimination and the high degree of our
polynomials. It is possible and will suffice, however, to {\it
bound} the number of solutions using the following result due to
Buchberger \cite{gedd92}:

\begin{theorem}
\la{buch} Let $G$ be a Gr\"{o}bner basis for $<F>$ (the polynomial
ideal generated by $F$) with respect to a given ordering of terms,
and let $H$ denote the set of leading terms of the elements of $G$
with respect to the chosen term ordering. Then the system of
equations corresponding to $F$ has finitely many solutions if and
only if for every indeterminate $x$ in $F$ there is a natural
number $m$ such that $x^m \in H$.
\end{theorem}
The key to using this result is that we may use an ordering of
terms based on {\it total degree} (i.e. a non-elimination
ordering) for which the computational complexity of Buchberger's
algorithm for Gr\"{o}bner bases is much lower. Unfortunately, even
in this setting a Gr\"{o}bner basis for $F$ cannot easily be
computed due to the extreme size of intermediate results produced
by the algorithm.

It would be highly desirable to apply modular homomorphisms in the
manner used in algorithms for factorization (e.g. so-called
Chinese remainder, or Hensel algorithms \cite{gedd92}) in the
present situation. This is not currently possible due to a number
of unresolved problems with the approach.  Nonetheless, it
provides a useful probabilistic experimental approach:  treat the
elements of $F$ as polynomials over a prime field $Z_p$ (rather
than the rationals), where $p$ is of modest size, and compute the
Gr\"{o}bner basis of $F$ $modulo$ $p$ over $Z_p$. For a {\it
single} prime, it is possible that the result so obtained may have
no useful relationship with the Gr\"{o}bner basis of $F$ over the
rationals. However, if the basis polynomials computed using a
large number of different primes all exhibit identical monomial
structures, it is extremely likely that they each represent a
distinct homomorphic image of the true Gr\"{o}bner basis of $F$.
The question of accurately computing the probability of success
for a specific series of primes remains an open problem. However,
the individual prime field computations are comparatively easy
since (unlike the rational case) no single coefficient may be
larger than the chosen prime.  This provides an experimental
"sampling" method which gives clues on how best to compute the
true result, and what that result will likely be.

We must also consider that if we were able to compute a
Gr\"{o}bner basis for $F$ over the rationals, we would derive
information on all solutions of the corresponding system {\it
including} those which were examined in the previous section (i.e.
for which $\Phi_{11} = 0$). It is possible to exclude those
solutions entirely by adding an additional constraint and
variable, \be \la{5.101} \phi_{11} z - 1 = 0 \ee to our equations
to produce the augmented system $\tilde F$. Still, only an actual
computation reveals whether this improves or worsens the
tractability of the problem. In our case, a large number (a few
thousand) prime field ``sample" computations (done using the GB
package of Faug\`ere \cite{fau94}, which is far more efficient
than the general-purpose Maple system) all suggested that the
addition of equation (\ref{5.101}) made the Gr\"{o}bner basis
calculation much more efficient. More importantly, once the
solutions examined in the previous section were in effect
discarded, only a {\it finite number} remained when Theorem
\ref{buch} is taken into account. With this in mind, it was
possible (and worthwhile) to compute the {\it true} Gr\"{o}bner
basis of $\tilde F$ over the rationals in the indeterminates $\x$,
$\y$, $\xc$, $\yc$, $r_{11}$, $z$ using a total degree ordering of
terms.  Since this basis contains polynomials with leading terms
\be \la{5.102} \x6, \y5, \xc5, \yc5, \phi_{11}^4, z5, \ee we may
conclude that there are only finitely many solutions for which
$\phi_{11}$, and hence $\Phi_{11}$ as well, is nonzero. (For this
last computation the latest and most efficient version of
Faug\`ere's GB package, known as FGB, was required.) It follows
that $\x$, $\y$, $\xc$, $\yc$, $\phi_{11}$ must be constants; it
remains only to show that this yields a contradiction.

Since $\phi_{11}$ must be constant (including the case in which
$\Phi_{11}=0$) it follows from (\ref{x_1}), (\ref{4.197}) that the
quantities $\pi\opi$, $\beta\obe$ and $\oal / \beta$ are all
constant as well. From the equation $\delta(\beta\obe) = 0$ we
obtain, in the variables $\x$, $\y$, $\phi_{11}$ given by
(\ref{x_1}), (\ref{4.197}), the side relation
\begin{eqnarray}
\nonumber
(7 \xc + 25 \y + 2) \phi_{11} + 374 \y \yc + 199 \xc \y \yc - 5 \y^2 \yc \\
\nonumber
+ 60 \x \xc \y + 60 \xc^2  \yc + 110 \x \y + 110 \xc \yc + 68 \x \xc  \\
\label{f.1} + 24 \x \xc^2 + 24 \xc^2 + 44 \x + 116 \xc + 132 =
0\,.
\end{eqnarray}
Next, from $\delta(\y) = 0$ we obtain the Pfaffian \be \la{f.2}
\delta \opi = - \opi ( \oal + \beta )\,. \ee Using this, along
with the previously determined Pfaffians, we then obtain from
$\delta(\pi \opi) = 0$ another side relation; on subtracting this
result from (\ref{f.1}) (and ignoring the possibility that $d_1 =
0$, which has already been considered) we obtain \be \la{f.3} \y (
\x + \yc + 4) + \phi_{11} = 0\,. \ee Finally, from $\ode(\oal /
\beta) = 0$ we obtain \be \la{f.4} \x(\x + 5\yc + 2) + 9\yc = 0\,.
\ee The collection of polynomials given by (\ref{f.1}),
(\ref{f.3}), (\ref{f.4}), (\ref{4.124b}) and their complex
conjugates has a Gr\"{o}bner basis (computed easily using Maple)
containing only the polynomial $1$. This is equivalent to showing
that there exists a combination of these polynomials which equals
$1$, and hence that they cannot vanish simultaneously (see
\cite{gedd92}); i.e., the associated system of equations has {\it
no} solutions. This completes the proof.

\section{Conclusion}
\la{sec-5} In completing the proof of Theorem \ref{teo-main}, we
have fully solved Hadamard's problem for the scalar wave equation
in the case of Petrov type III space-times. Essential to our proof
were use of the six-index necessary condition obtained by Rinke
and W\"unsch \cite{rink81}, and separate analyses (and different
ideal-theoretic tools) for the cases $\Phi_{11} = 0$ and
$\Phi_{11} \neq 0$. To complete the proof of the conjecture stated
in the Introduction it remains to consider the space-times of
Petrov types I and II.  A partial result for type II has been
obtained by Carminati, Czapor, McLenaghan and Williams
\cite{car91}.  However, it is not yet clear whether the
complicated equations which arise from conditions III, V, and VII
can be solved by the method used in the present paper.

The authors would like to thank J. C. Faug\`ere for his assistance
with the FGB package. This work was supported in part by the
Natural Sciences and Engineering Research Council of Canada in the
form of individual Research Grants (S. R. Czapor and R. G.
McLenaghan).

\bibliography{conf}
\bibliographystyle{plain}

\end{document}